\documentclass{article}
\usepackage{natbib}
\usepackage{graphicx} 
\usepackage[margin=1in]{geometry}
\usepackage{adjustbox}
\usepackage{setspace}
\usepackage{indentfirst}
\usepackage{multirow}
\usepackage{array}
\usepackage{amsmath}
\usepackage{amssymb}
\usepackage{bbm}
\usepackage{booktabs}
\usepackage{caption} 
\usepackage{enumerate}
\usepackage{enumitem}
\usepackage{xr-hyper}
\usepackage{threeparttable}
\usepackage{tablefootnote}
\usepackage{kpfonts}
\usepackage{bigints}
\makeatletter
\newcommand*{\addFileDependency}[1]{
  \typeout{(#1)}
  \@addtofilelist{#1}
  \IfFileExists{#1}{}{\typeout{No file #1.}}
}
\makeatother

\newcommand*{\myexternaldocument}[1]{%
    \externaldocument{#1}%
    \addFileDependency{#1.tex}%
    \addFileDependency{#1.aux}%
}

\myexternaldocument{./main_supp}

\title{Compatible Imputation for Hierarchical Linear Models with Incomplete Data: Interaction Effects of Continuous and Categorical Covariates MAR}
\author{Dongho Shin, Yongyun Shin}
\date{}
\begin{document}

\maketitle
\doublespacing
\begin{abstract}
This article focuses on Bayesian estimation of a hierarchical linear model (HLM) from incomplete data assumed missing at random where continuous covariates C and discrete categorical covariates $D$ have interaction effects on a continuous response $R$. Given small sample sizes, maximum likelihood estimation is suboptimal, and existing Gibbs samplers are based on a Bayesian joint distribution compatible with the HLM, but impute missing values of $C$ and the underlying latent continuous variables $D^*$ of $D$ by a Metropolis algorithm via proposal normal densities having constant variances while the target conditional distributions of $C$ and $D$ have nonconstant variances. Therefore, the samplers are neither guaranteed to be compatible with the joint distribution nor ensured to always produce unbiased estimation of the HLM. We assume a Bayesian joint distribution of parameters and partially observed variables, including correlated categorical $D$, and introduce a compatible Gibbs sampler that draws parameters and missing values directly from the exact posterior distributions. We apply our sampler to incompletely observed longitudinal data from the small number of patient-physician encounters during office visits, and compare our estimators with those of existing methods by simulation. \\ 
$Keywords$: Compatibility, Correlated Categorical Variables, Exact posterior distributions, Gibbs sampler, Multiple Imputation, Missing at random.
\end{abstract}

\pagebreak
\section{Introduction}
Medical researchers frequently encounter analysis of hierarchical or multilevel data collected, for example, from patients nested within physicians or clinics, residents within neighborhoods, and repeated measurements within patients. A hierarchical linear model (HLM), also known as a variance component, multilevel or linear mixed model, is appropriate to analyze such data with a nested structure \citep{raudenbush2002hierarchical, goldstein2003multilevel, snijders2011multilevel, verbeke2013generalized}. Missing data at either level of the hierarchy commonly arises, making it difficult to achieve efficient and unbiased estimation of the HLM. Complete-case analysis is inefficient and may produce substantially biased estimates \citep{little2002bayes}. In particular, it is costly to drop a cluster with cluster-level missing data in a HLM as all units nested within the cluster also has to be dropped. The resulting estimates are inefficient and may be substantially biased.

Multiple imputation (MI) \citep{rubin1987, rubin1996multiple} is a popular approach to handling missing data. In multilevel models, joint modeling approach and MI by fully conditional specification (FCS) are commonly used to impute hierarchical missing data assumed missing at random \citep{rubin1976inference}. Joint modeling approach estimates a joint distribution of partly observed variables, including the response variable, and missing values are imputed from their predictive distribution given observed data \citep{schafer2002computational, shin2007just,enders2018fully}. FCS estimates a  univariate model for each partially observed variable conditional on all other variables, and imputes the missing values of the variable from the fitted conditional model \citep{raghunathan2001multivariate, van2007multiple, van2011mice, van2018flexible}. These approaches were shown to work well when the joint distribution of incompletely observed variables are reasonably normally distributed \citep{liu2014stationary, bartlett2015multiple, enders2016multilevel}. 

With interaction or other nonlinear effects of partially observed covariates in a HLM, however, the joint distribution of incompletely observed variables is not multivariate normal even if the factorized conditional distributions may be normally distributed \citep{hughes2014joint,sullivan2017multiple, kim2018multiple, shin2024maximum, shin2024bayesian}. MI of missing data under the joint normality assumption may produce substantially biased estimates \citep{enders2020model, shin2024maximum}.

Our focus in this paper is on a mixture of correlated nominal categorical covariates $D$ and continuous covariates $C$ having interaction effects on a continuous response variable $R$ in a HLM. FCS imputes each missing value from a univariate logistic, multinomial logistic or linear mixed regression model\citep{van2007multiple, van2018flexible}. Except for a simple bivariate distribution of binary and continuous variables \citep{liu2014stationary, efron1975efficiency}, the chained regressions are not compatible with a Bayesian joint distribution \citep{hughes2014joint, bartlett2015multiple, liu2014stationary, enders2020model}. 

In this paper, we extend the general location model \citep{olkin1961multivariate, little1985maximum} where a vector of correlated $q$ nominal categorical variables $D$ is expressed as a multinomial random variable in a $q$-way contingency table and continuous variables are assumed multivariate normal conditional on discrete categorical $D$. \citet{little1985maximum} implemented maximum likelihood (ML) estimation of a single-level general location model to handle a mixture of categorical and continuous missing data by the EM algorithm \citep{dempster1977maximum}. In this paper, we extend this approach to a HLM where $p$ continuous covariates $C$ and $q$ categorical covariates $D$ have interaction effects on a continuous response $R$.

To handle missing values given incompletely observed categorical $D$ and continuous $(R,C)$ in the HLM, an efficient approach assumes joint normality of $(R,C)$ and latent continuous $D^*$ underlying $D$ \citep{carpenterandkenward, hughes2014joint, bernaards2007robustness, sullivan2017multiple}. \citet{goldstein2014fitting} imputed missing values of covariates having nonlinear effects on a continuous or binary outcome by the joint normal approach via a Gibbs sampler and a Metropolis algorithm \citep{hastings1970monte}. \citet{enders2020model} extended the Gibbs sampler to handling categorical and continuous predictors MAR having nonlinear effects in two- and three-level models, and illustrated how FCS by MICE \citep{van2011mice} may lead to an imputation model incompatible with an analytic hierarchical model to produce biased estimates. Based on the underlying multivariate normality, the joint normal approach is unable to estimate nonlinear effects of covariates well \citep{hughes2014joint, erler2019bayesian}. In the presence of the non-linearities, the Bayesian joint distribution based on the underlying joint normality assumption is not guaranteed to be compatible with the HLM \citep{kim2015evaluating, bartlett2015multiple, goldstein2014fitting,kim2018multiple, enders2020model}. The joint modeling approach \citep{schafer2002computational, shin2024maximum} has yet to be extended to the HLM involving the interaction effects of correlated elements of partially observed $C$ and $D$. 

We contribute to the literature by introducing a Gibbs sampler that is based on a Bayesian joint distribution expressing the joint distribution of $R, C, D$ given parameters and a reasonably assumed prior distribution of parameters \citep{schafer2002computational, enders2020model, shin2024bayesian}. Our Gibbs sampler imputes parameters and missing values of the correlated categorical $D$ and continuous $C$ having interaction effects on a continuous $R$ in the HLM from their exact posterior distributions. See \citet{shin2024bayesian} for a Gibbs sampler that estimates a HLM with the interaction effects of continuous covariates $C$ on a continuous response $R$ by imputing missing values and parameters from their exact posterior distributions. Therefore, our Gibbs sampler is guaranteed to be compatible with the Bayesian joint distribution and, thus, produce unbiased estimation of the HLM while existing Gibbs samplers ensures  neither their compatibility with the HLM and HLM nor unbiased estimation of the HLM.

Our motivating application is analysis of racially discordant medical interactions between patients and physicians from different racial groups. During office visits, these interactions were videotaped and coded to produce four repeated measurements of a positive valence score rating the goodness of physicians' interaction with patients by their facial expression and communication behavior. The focus is on the main and interaction effects of physician's implicit prejudice and the income levels of patients on the valence outcome in a HLM where repeated measurements of the outcome is nested within physician-patient encounters. Four key challenges motivated the development of our Bayesian approach. First, the valence score, physician's implicit prejudice, and patient income include 20\%, 16\% and 11\% of their values missing, respectively, at the encounter or cluster level. Second, the effects of the partially observed implicit prejudice and income is hypothesized to produce an interaction effect, thereby making unbiased and efficient estimation challenging. Furthermore, the interaction effect of a mixture of a continuous variable and a binary variable adds extra complexity to achieving the compatibility between our Gibbs sampler and the HLM. Finally, due to COVID-related restrictions, only 37 encounters were recorded, involving a small number of 6 physicians and 37 patients. The small sample sizes pose a formidable challenge to efficient and unbiased estimation of the HLM.

The rest of the chapter is organized as follows. In section 2, we introduce our analytic HLM. In section 3, we present our Gibbs sampler based on the exact posterior distributions of missing covariates. In section 4, we assess the performance of our method by comparing our estimators with those by existing methods via simulation. In section 5, we apply our method to analyze the racially discordant patient-physician encounter data. Lastly, in section 6, we discuss some limitations and future extensions of our approach. 
\clearpage
\section{Model}
Our interest focuses on a two-level hierarchical linear model (HLM)
\begin{flalign} \label{eq:4model}
Y_{ij}=\beta_{0}+\beta^T_C C_{j}+\beta^T_D D_{j}+\beta^T_X X_{ij}+D^T_{j}\otimes C^T_{j} \beta_{CD}+u_{j}+e_{ij} 
\end{flalign}
\noindent where $Y_{ij}$ is the outcome variable, $C_j=[C_{1j} ... C_{pj}]^T$ is a $p$-by-$1$ vector of cluster-level continuous covariates having main effects $\beta_C$, $D_j=[D_{1j} ... D_{qj}]^T$ is a $q$-by-$1$ vector of cluster-level categorical covariates having main effects $\beta_D$, and $X_{ij}$ is a vector of fully observed level-1 covariates having main effects $\beta_X$. In addition, $\beta_{CD}$ is the $(p\times q)$-by-$1$ interaction effects of $C_{j}$ and $D_{j}$. Alternatively, we express $D^T_{j}\otimes C^T_{j} \beta_{CD}=C^T_{j}B_{CD}D_j 
$ for a $p\times q$ matrix $B_{CD}=[\beta_{CD1} \beta_{CD2} \cdots \beta_{CDq}]$ and $\beta_{CD}=vec(B_{CD})=[\beta^T_{CD1} \cdots \beta^T_{CDq}]^T$ where kronecker product $A\otimes B$ multiplies matrix $B$ to each element of matrix $A$ and $vec(A)$ stacks the columns of $A$ up or vectorizes the columns. Lastly, a level-2 unit-specific random effect $u_j\sim \mathcal{N}(0,\tau)$ and a level-1 unit specific random effect $e_{ij} \sim N(0,\sigma^2)$ are independent, and a level-1 unit $i$ is nested within a level-2 cluster $j$ for $i=1,\cdots,n_j$ and $j=1,\cdots,J$. For some elements of $D_j$ that have only main effects, we set the corresponding components of $\beta_{CD}$ to zero. Here, covariates $C_{j}, D_j$ and the outcome $Y_{ij}$ may be partially observed. 

\section{Compatible Gibbs Sampler}
This section explains our Gibbs sampler based on exact posterior distributions. We assume a Bayesian joint distribution of $\mathbf{Y}=(Y_{11},Y_{12},\cdots,Y_{n_JJ}), \mathbf{C}=(C_1.C_2,\cdots,C_J), \mathbf{D}=(D_1.D_2,\cdots,D_J)$ and $\theta$ given $\mathbf{X}=(X_{11},X_{12},\cdots,X_{n_JJ})$
\begin{align} \label{eq:4joint}
    f(\mathbf{Y},\mathbf{C},\mathbf{D},\theta|\mathbf{X})= 
    \prod_{j=1}^J \prod_{i=1}^{n_j}f(Y_{ij}|C_{j},D_j,X_{ij},u_j,\boldsymbol{\beta}, \sigma^2)f(u_j|\tau)f(C_{j}|D_j,\alpha,T)f(D_j|\pi)p(\theta),
\end{align}
\noindent Here, $f(Y_{ij}|C_{j},D_j, X_{ij},u_j,\boldsymbol{\beta},\sigma^2)$ and $f(u_j|\tau)$ are the normal densities from the HLM (1) where $\boldsymbol{\beta}=(\beta_0, \beta_C,$ \\ $\beta_D, \beta_{X}, \beta_{CD})$ for $\beta_{CD}=[\beta^T_{CD1} \cdots \beta^T_{CDp}]^T$. To handle missing values of $C_j$ efficiently, we assume   
\begin{align}\label{eq:4covariate}
f(C_{j}|D_j) \sim N\left[ W\alpha = \left(I_{p} \otimes [1\; D^T_{j}]\right) \alpha, T\right]
\end{align}
\noindent for a vector of fixed effects $\alpha$, a $p\times p$ identity matrix $I_{p}$ and a $p\times p$ variance-covariance matrix $T$. We explain $f(D_j|\pi)$ in detail below. The prior $p(\theta)$ of $\theta=(\boldsymbol{\beta}, \tau, \sigma^2, \alpha,T,\pi)$ is specified in the Gibbs sampler steps
below.

\subsection{Exact posterior Distributions of Continuous Covariates}
Let $p(A|\cdot)$ denote the posterior, or exact posterior, distribution of A given all other unknowns. To find the key conditional distribution $p(C_{kj}|\cdot)$ for a missing $C_{kj}$ of $C_j$, we first derive a bivariate normal conditional distribution of $Y_{ij}$ and $C_{kj}$ from Equations (\ref{eq:4model}) and (\ref{eq:4covariate}) 
\begin{gather} \label{eq:4bivariate}
    f(Y_{ij},C_{kj}|C_{(-k)j},D_j, X_{ij},u_j,\boldsymbol{\beta}, \sigma^2,\alpha,T) \sim N\left(\begin{bmatrix}
 \mu_{1ij}+\mu_{2ij}M_{k|(-k)}\\ M_{k|(-k)}
\end{bmatrix}, \begin{bmatrix}
\mu_{2ij}^2T_{k|(-k)}+\sigma^2 & \mu_{2ij}T_{k|(-k)}  \\
\mu_{2ij}T_{k|(-k)} & T_{k|(-k)}  \\
    \end{bmatrix}\right)
    \intertext{for $C_{(-k)j}=(C_{1j},...,C_{(k-1)j},C_{(k+1)j},...,C_{pj})$, $M_{k|(-k)}=E(C_{kj} | C_{(-k)j})$ and $T_{k|(-k)}=var(C_{kj} | C_{(-k)j})$. The conditional mean $E(Y_{ij}|C_{(-k)j}, D_j, X_{ij}, u_j)$ has two parts, $\mu_{1ij}$ excluding and $\mu_{2ij}$ including $C_{kj}$, to facilitate derivation of $p(C_{kj}|\cdot)$:}
    \begin{aligned}
&\mu_{1ij}=\beta_{0}+\beta^T_{C(-k)}C_{(-k)j}+\beta^T_XX_{ij}+C_{(-k)j}B_{CD(-r)}D_j+u_j, \\ \nonumber
&\mu_{2ij}
=\left(\beta_{C_k}+B_{CDr}D_j\right)
\end{aligned}
\end{gather}
where $\beta_C$ is partitioned into the coefficients $\beta_{Ck}$ of $C_{kj}$ and others $\beta_{C(-k)}$, $B_{CDr}$ is the $r$th row of $B_{CD}$, and $B_{CD(-r)}$ is a $(p-1)\times q$ matrix after deleting $B_{CDr}$ from $B_{CD}$.

The joint distribution (\ref{eq:4joint}) then implies
\begin{align} \label{eq:4exact}
p(C_{kj}|\cdot) & \propto \prod_{i=1}^{n_j}f(Y_{ij}|C_{j},D_j,X_{ij},u_j,\boldsymbol{\beta}, \sigma^2) \times f(C_{kj}|C_{(-k)j},D_j,x_{2j}, \alpha, T) \sim N(\Tilde{M}_{kj}, \Delta_{kj}^{-1})
\end{align}
where $\Tilde{M}_{kj}=M_{k|(-k)}+\Delta^{-1}_{kj}\sigma^{-2}\mu_{2ij}\sum_{i}^{n_j}\left(Y_{ij}-(\mu_{1ij}+\mu_{2ij}M_{k|(-k)})\right)$ and $\Delta_{kj}=T^{-1}_{k|(-k)}+n_j\mu_{2ij}^2\sigma^{-2}.$

\subsection{Exact Posterior Distribution of Correlated Categorical Covariates}
\subsubsection{Notations}
Correlated $q$ nominal categorical variables $D_j=(D_{1j},...,D_{qj})$ of cluster $j$ are cross-classified into a $q$-way contingency table $\mathcal{T}$ with $\mathcal{C}$ cells where the $k$th categorical variable has $L_k$ level, and $\mathcal{C}=\prod_{k=1}^qL_k$. We partition $D_j=(D_{obsj}, D_{misj})$ for observed $D_{obsj}$ and missing $D_{misj}$. Observed $D_{obsj}$ determines a set of cells to which cluster j may belong. Following the notations from \citet{little1985maximum}, we denote the set $S_j$. For example, consider 3 binary variables $D_j=(D_{1j},D_{2j},D_{3j})$, resulting in $\mathcal{T}$ with $2^3=8$ cells. Suppose that $D_{misj}=(D_{1j},D_{2j})$ and $D_{obsj}=D_{3j}=1$ for cluster j. Consequently, $S_j$ consists of four cells: $D_j=(1,1,1), (1,0,1), (0,1,1)$ and $(0,0,1)$. 

Let $W_j$ be a $\mathcal{C} \times 1$ random vector from $Multinomial(1, \pi_1,\cdots,\pi_{\mathcal{C}})$ where $\pi_{d}=P(W_j=E_d)$ for a $\mathcal{C} \times 1$ vector $E_d$ having a single 1 in cell $d$ and all others equal to zero. Note that $W_j$ is missing unless all elements of $D_j$ are fully observed. For the cluster with $D_{obsj}=D_{3j}$, for example, $d=1,2,3,4$ refer to cells $D_{misj}=(D_{1j},D_{2j})=(1,1), (1,0), (0,1)$ and $(0,0)$, respectively, given $D_{3j}=1$; we denote  $D_{misj}=\mathbf{d}_{dj}$ for $\mathbf{d}_{1j}=(1,1), \mathbf{d}_{2j}=(1,0), \mathbf{d}_{3j}=(0,1)$ and $\mathbf{d}_{4j}=(0,0)$ with cell-d probability $P(W_{misj}=E_d | D_{obsj})$ for $E_1=(1,0,0,0), E_2=(0,1,0,0), E_3=(0,0,1,0), and E_1=(0,0,0,1)$, respectively.

\subsubsection{exact posterior distribution}
Let $D_{misj}$ be a $q_j$ by $1$ vector for $q_j\le q$. There are $\mathcal{C}_j=\prod_{k=1}^{q_j} L_k$ cells in $S_j$. We now describe the key conditional probability mass function $p(D_{mis,j}|\cdot)$
\begin{align*}
p(D_{mis,j}|\cdot) \propto h(D_{mis,j})= 
\prod_{i=1}^{n_j} f(Y_{ij}|C_{j}, D_j, X_{ij},u_j,\boldsymbol{\beta}, \sigma^2)f(C_{j}|D_j,\alpha,T)f(D_{j}|\pi),
\end{align*}
where the posterior probability $\pi_{dj}=P(D_{misj}=\mathbf{d}_{dj}|\cdot)$ for a cell $d\in S_j$ is computed by
\begin{align*}
    \pi_{dj} = \frac{h(\mathbf{d}_{dj})}{\sum_{d' \in S_j} h(\mathbf{d}_{d'j})}
\end{align*}

\noindent We find a $q_j$-way contingency table $\mathcal{T}_j$ with $\mathcal{C}_j$ cells and cell-d probability $P(W_{misj}=E_d)=\pi_{dj}$: 
\begin{align} \label{eq:4Wmultinomial}
p(W_{misj}|\cdot) \sim Multinomial(1,\pi_{1j},\cdots,\pi_{\mathcal{C}_jj}).                \end{align}      
\clearpage

\subsection{Gibbs Sampler Steps Based on Exact Posterior Distributions}
We partition complete data $\mathbf{Y}=(\mathbf{Y}_{obs},\mathbf{Y}_{mis})$, $\mathbf{C}=(\mathbf{C}_{obs},\mathbf{C}_{mis})$, and $\mathbf{D}=(\mathbf{D}_{obs},\mathbf{D}_{mis})$ into observed \\
$(\mathbf{Y}_{obs}, \mathbf{C}_{obs},\mathbf{D}_{obs})$ and missing $(\mathbf{Y}_{mis},\mathbf{C}_{mis},\mathbf{D}_{mis})$. Bayesian joint distribution (2) implies the conditional distribution of $u_j$ and posterior distributions of $\theta$:
\begin{align*}
 &p(u_j|\cdot) \propto \prod_{i=1}^{n_j}f(Y_{ij}|C_{j}, X_{ij}, u_j, \boldsymbol{\beta}, \sigma^2)f(u_j|\tau), \\
&p(\tau|\cdot) \propto \prod_{i=1}^{n_j}f(u_j|\tau)p(\tau),\\
&p(\boldsymbol{\beta}|\cdot) \propto \prod_{j=1}^{J}\prod_{i=1}^{n_j}f(Y_{ij}|C_{j}, X_{ij},u_j, \boldsymbol{\beta}, \sigma^2)f(\boldsymbol{\beta}),\\
&p(\sigma^2|\cdot) \propto \prod_{j=1}^{J}\prod_{i=1}^{n_j}f(Y_{ij}|C_{j},X_{ij},u_j, \boldsymbol{\beta}, \sigma^2)p(\sigma^2),\\
&p(\alpha|\cdot) \propto \prod_{j=1}^{J}f(C_{j}|\alpha,T)p(\alpha),\\
&p(T|\cdot) \propto \prod_{j=1}^{J}f(C_{j}|\alpha,T)p(T)  \\
&p(\pi|\cdot) \propto \prod_{j=1}^{J}f(W_j|\pi)p(\pi)
\end{align*}

\noindent where we assume priors: inverse gamma $p(\tau)\sim IG(\alpha_0=1, \beta_0=0.5)$ and $p(\sigma^2)\sim IG(\alpha_0=1, \beta_0=0.5)$, an inverse wishart $p(T)\sim IW(V_0,S_0^{-1})$ with $V_0=p+2$ and $S_0=\hat{T}$ (estimated variance-covariance matrix using complete cases), noninformative $p(\beta)=p(\alpha)=1$, and $p(\pi)\sim Dirichlet(a_i=1 \forall i=1,...,\mathcal{C})$, following \citet{schafer2002computational}, \citet{hoff2009first}, \citet{si2013nonparametric} and \citet{enders2020model}. With a univariate binary $D_j$, the dirichlet prior becomes $beta(1,1)$.

At cycle $t$ given parameters $\theta=\theta^{(t-1)}$ and completed data $\mathbf{Y}=(\mathbf{Y}_{obs},\mathbf{Y}^{(t-1)}_{mis})$, $\mathbf{C}=(\mathbf{C}_{obs},\mathbf{C}^{(t-1)}_{mis})$, and $\mathbf{D}=(\mathbf{D}_{obs},\mathbf{D}^{(t-1)}_{mis})$ from cycle $t-1$, we draw parameters and missing values by the following Gibbs sampler steps.

\begin{enumerate}[label=Step \arabic*:,leftmargin=*]
\item we sample $u^{(t)}_j$ from 
\begin{align*}
  p(u_j|\mathbf{Y}=\mathbf{Y}^{(t-1)},\theta=\theta^{(t-1)}) \sim N\left(\Delta_j^{-1}\sigma^{-2} \sum_{i=1}^{n_j}(Y_{ij}-\mathbf{X}^T_{ij}\boldsymbol{\beta}), \Delta_j^{-1}\right),
\end{align*}     
\noindent where $\mathbf{X}^T_{ij}=\left[1\; C^T_{j}\; D^T_{j}\; X^T_{ij}\; D^T_j\otimes C^T_j\right]$, $\boldsymbol{\beta}=\left[\beta_0\; \beta^T_C\; \beta^T_D\; \beta^T_{X}\; \beta^T_{CD}\right]^T$ and $\Delta_j=n_j\sigma^{-2}+\tau^{-1}$.  

\item Draw $\tau^{(t)}$ from
\begin{align*}
    p(\tau | \mathbf{u}=\mathbf{u}^{(t)}, \theta=\theta^{(t-1)}) \sim IG\left(\frac{J}{2}+\alpha_0, \Bigr[\frac{\sum_{i=1}^{n_j}u_j^2}{2}+\frac{1}{\beta_0}\Bigr]^{-1}\right).
\end{align*} \par

\item Draw $\boldsymbol{\beta}^{(t)}$ from
\begin{align*}
    p(\boldsymbol{\beta}|\mathbf{Y}=\mathbf{Y}^{(t-1)} \mathbf{u}=\mathbf{u}^{(t)}, \sigma^{2 (t-1)}) \sim N\left(\Bigr(\sum_{j=1}^{J}\sum_{i=1}^{n_j}\mathbf{X}_{ij}\mathbf{X}_{ij}^T\Bigr)^{-1}\sum_{j=1}^{J}\sum_{i=1}^{n_j}\mathbf{X}_{ij}(Y_{ij}-u_j),\sigma^2\Bigr(\sum_{j=1}^{J}\sum_{i=1}^{n_j}\mathbf{X}_{ij}\mathbf{X}_{ij}^T\Bigr)^{-1}\right). 
\end{align*} 

\item Draw $\sigma^{(t)}$ from 
\begin{align*}
    p(\sigma^2|\mathbf{Y}=\mathbf{Y}^{(t-1)}, \mathbf{u}=\mathbf{u}^{(t)}, \boldsymbol{\beta}=\boldsymbol{\beta}^{(t)}) \sim IG\left(\frac{N}{2}+\alpha_0, \Bigr[\frac{\sum_{j=1}^{J}\sum_{i=1}^{n_j}e_{ij}^2}{2}+\frac{1}{\beta_0}\Bigr]^{-1}\right),
\end{align*} 
where $N=\sum_{j=1}^J n_j$.

\item For a missing outcome $Y_{ij}$, impute $e^{(t)}_{ij}$ from $p(e_{ij}|\sigma^2=\sigma^{2 (t)}) \sim N(0,\sigma^2)$ and set $Y^{(t)}_{ij}=\mathbf{X}^T_j\boldsymbol{\beta}^{(t)}+u^{(t)}_j+e^{(t)}_{ij}.$ 

\item Draw $\alpha^{(t)}$ from
\begin{align*}
    p(\alpha|\mathbf{C}=\mathbf{C}^{(t-1)},\mathbf{D}^{(t-1)}, T^{(t-1)}) \sim N\left(\Bigr(\sum_{j=1}^{J}W^TT^{-1}W\Bigr)^{-1}\sum_{j=1}^{J}W^TT^{-1}C_j, \Bigr(\sum_{j=1}^{J}W^TT^{-1}W\Bigr)^{-1}\right). 
\end{align*} 

\item Draw $T^{(t)}$ from
\begin{align*}
    p(T | \mathbf{C}=\mathbf{C}^{(t-1)},\mathbf{D}^{(t-1)}, \alpha^{(t)}) \sim IW\left(V_0+J, \left(S_0+\sum_{j=1}^{J}\left(\begin{bmatrix}C_j \\ x_{2j}\end{bmatrix}-W\alpha\right)\left(\begin{bmatrix}C_j \\ x_{2j}\end{bmatrix}-W\alpha\right)^T\right)^{-1}\right). 
\end{align*} 

\item Draw $\pi^{(t)}$ from 
\begin{align*}
p(\pi|\cdot) \sim Dirichlet(a_1+\sum_{j=1}^{J} I(W_j=E_1),...,a_\mathcal{C}+\sum_{j=1}^{J} I(W_j=E_{\mathcal{C}}), 
\end{align*} 
for an indicator function $I(B)=1$ if condition $B$ is true and 0 otherwise.

\item For each $C_{kj}$ missing, define
\begin{align*}
    C^{(t)}_{(-k)j}=\left(C^{(t)}_1,...,C^{(t)}_{k-1}, C^{(t-1)}_{k+1},...,C^{(t-1)}_{p}\right) 
\end{align*}
that consists of $(k-1)$ observed or imputed missing values at cycle $t$ and $(p-k)$ observed or imputed values at cycle $t-1$. We compose a bivariate distribution (\ref{eq:4bivariate})
\begin{align*}
    f(Y_{ij},C_{kj}|\mathbf{Y}=\mathbf{Y}^{(t)}, C_j=C^{(t)}_{(-k)j}, D_j=D^{(t-1)}_j,u_j=u^{(t)}_j,\theta=\theta^{(t)})
\end{align*}
to draw $C_{kj}$ from the implied conditional distribution (\ref{eq:4exact})
\begin{align*}
    p(C_{kj}| \mathbf{Y}=\mathbf{Y}^{(t)}, C_j=C^{(t)}_{(-k)j}, D_j=D^{(t-1)}_j, u_j=u^{(t)}_j,\theta=\theta^{(t)}).
\end{align*}

\item Draw $W_{misj}$ from the conditional distribution (\ref{eq:4Wmultinomial}) :
\begin{align*}
     f(W_{misj}|\mathbf{Y}=\mathbf{Y}^{(t)}, C_j=C^{(t)}_j, D_j=D^{(t-1)}_j, u_j=u^{(t)}_j,\theta=\theta^{(t)})
\end{align*}
and, then, translate the $W_{misj}$ into the imputed values $D_{misj}$.
\end{enumerate}

\section{Simulation Study}

We now evaluate our compatible Gibbs sampler based on exact posterior distributions (GSExact) by simulation studies. The simulated data from a HLM (\ref{eq:4model}) will resembles real data that we analyze in the next section in terms of correlations, sample sizes, and missing rate. We then compare our estimators with those by 1) the lme4 package in R \citep{douglas2015fitting} that estimates maximum likelihood estimates given complete data (CDML) and 2) software Blimp \citep{blimp} that implements the Gibbs Sampler of \citet{enders2020model}. We do not consider FCS that was shown to be incompatible given the non-linearities of interest in this paper (\citealp{liu2014stationary}; \citealp{bartlett2015multiple}; \citealp{enders2016multilevel}; \citealp{enders2020model}; \citealp{kim2015evaluating}; \citealp{kim2018multiple}). Because CDML is provided with complete data while others are based on data MAR, a good method will produce estimates near those by CDML.

We consider two cases where $n_j=4$ units are nested within each of 1) $J=36$ clusters or a small sample and 2) $J=200$ clusters or a large sample. We focus on the analytic HLM we analyzed in the next section where one continuous covariate MAR and one binary covariate MAR have main and interaction effects. We validate the correct execution of our R code that implements our sampler given the large sample and compare our estimators with the competing ones given the small and large samples. 

We simulate sequentially: 1) $D_j\sim Bernoulli(\pi_d=0.3)$; 2) $C_{2j}\sim N(-0.5+D_j, 1)$; 3) $C_{1j} \sim N(0.5-0.5X_j+1.2D_j, 1)$; 4) $Y_{ij}\sim N(\beta_0+\beta_1C_{1j}+\beta_2D_j+\beta_3C_{2j}+\beta_4C_{1j}D_j, \tau+\sigma^2)$ for $\tau=4$ and $\sigma^2=16$. The simulated coefficients of the HLM are all equal to $\beta_0=\beta_1=\beta_2=\beta_3=\beta_4=1$ for easy comparison.

We simulate the missing rate of $Y_{ij}$, $C_{1j}$ and $D_j$ by a MAR mechanism that depends on fully known $C_{2j}$.
\begin{align} \label{eq:4missingrate}
logit(p) \sim N(c_0+c_1C_{2j}, \delta) 
\end{align}
\noindent and set a missing value by Bernoulli(p). We manipulate $c_0$, $c_1$ and $\delta$  to simulate the missing rates of real data in the next section. We set a missing value of $Y_{ij}$ given $c_0=-1.9, c_1=0.1$ and $\delta=1$; $C_{1j}$ given $c_0=-2.2, c_1=-1.5$ and $\delta=0$; and $D_j$ given $c_0=-2.0, c_1=1.5$ and $\delta=0$. Consequently, each variable is missing about 20\% of the values.

We repeated simulating data and estimating the HLM 1000 times to compute the \% bias, average estimated standard error (ASE), empirical estimate of the true standard error (ESE) over samples and 95\% coverage probability (coverage) of each estimator. Both Blimp and our sampler are based on 2500 burn-in and 2500 post-burn iterations.

\begin{table}[htbp]
  \caption{Estimated biases, ASEs, ESEs and coverages from the large sample simulation $(n_j=4, J=200)$.}
\begin{adjustbox}{max width=\textwidth}
\renewcommand{\arraystretch}{1.5}
    \begin{tabular}{cccccccccc}
\toprule
  & \multicolumn{3}{c}{CDML} & \multicolumn{3}{c}{GSExact} & \multicolumn{3}{c}{Blimp} \\
\cmidrule(lr){2-4}\cmidrule(lr){5-7}\cmidrule(lr){8-10}
 Simulated & \%Bias(ASE) & ESE & Coverage & \%Bias(ASE) & ESE & Coverage  & \%Bias(ASE) & ESE & Coverage \\ \midrule

 $\tau$=4 & 1.6 (-) & 0.84 & -& -2.6 (0.99) & 1.00  & 0.94 & 3.9 (1.04) & 1.01 & 0.95 \\

 $\sigma^2$=16 & -0.4 (-) & 0.93 & - & 0.4 (1.06) & 1.07 & 0.94 & 0.2 (1.05) & 1.06  & 0.94  \\
 
 $\beta_{0}$=1 & -1.1 (0.29) & 0.30 & 0.94 & -2.1 (0.33) & 0.35 & 0.94 & -0.9 (0.33) & 0.34 & 0.95 \\

 $\beta_{1}$=1 & 0.7 (0.24) & 0.24 & 0.94 & -0.9 (0.28) & 0.29 & 0.94 & 0.5 (0.28) & 0.29 & 0.94 \\

 $\beta_{2}$=1 & 0.8 (0.74) & 0.75 & 0.95 & -2.2 (0.88) & 0.91 & 0.93 & 3.4 (0.91) & 0.93 & 0.94 \\

 $\beta_{3}$=1 & 0.8 (0.23)  & 0.23 & 0.95 & 1.4 (0.26) & 0.26  & 0.94 & 0.7 (0.26) & 0.25 & 0.95 \\

 $\beta_{4}$=1 & -1.0 (0.40)  & 0.41 & 0.95 & -0.9 (0.48) & 0.49  & 0.94 & -2.1 (0.49) & 0.50 & 0.94 \\
 
\bottomrule
\end{tabular}%
\end{adjustbox}
\label{catej=200}%
\end{table}%

Table \ref{catej=200} summarizes the simulation results for the large sample scenario. Both GSExact and Blimp produce estimates close to those of CDML. All three approaches produce estimates very accurate and precise estimates with biases $<$ 3\% except for the estimated $\tau$ (3.9\% biased) and effect $\beta_{2}$ of $C_{1j}$ (3.4\% biased) by Blimp. ASEs are close to ESEs, and coverage probabilities near the nominal 0.95. CDML by the lme4 package does not produce variances associated with variance estimates. The standard errors from GSExact and Blimp are comparatively inflated, reflecting additional uncertainty resulting from missing data.  

\begin{table}[htbp]
  \caption{Estimated biases, ASEs, ESEs and coverages from the small sample simulation $(n_j=4, J=36)$.}
\begin{adjustbox}{max width=\textwidth}
\renewcommand{\arraystretch}{1.5}
    \begin{tabular}{cccccccccc}
\toprule
  & \multicolumn{3}{c}{CDML} & \multicolumn{3}{c}{GSExact} & \multicolumn{3}{c}{Blimp} \\
\cmidrule(lr){2-4}\cmidrule(lr){5-7}\cmidrule(lr){8-10}
 Simulated & \%Bias(ASE) & ESE & Coverage & \%Bias(ASE) & ESE & Coverage  & \%Bias(ASE) & ESE & Coverage \\ \midrule

 $\tau$=4 & 0.4 (-) & 2.09 & -& -7.4 (2.23) & 1.80  & 0.96 & 18.9 (3.14) & 2.43 & 0.97 \\

 $\sigma^2$=16 & -0.7 (-) & 2.21 & - & 0.8 (2.47) & 2.39 & 0.94 & 1.4 (2.56) & 2.41  & 0.95  \\
 
 $\beta_{0}$=1 & -1.0 (0.71) & 0.73 & 0.94 & 0.9 (0.83) & 0.84 & 0.95 & 3.0 (0.89) & 0.81 & 0.96 \\

 $\beta_{1}$=1 & -1.5 (0.58) & 0.58 & 0.95 & -1.3 (0.74) & 0.73 & 0.95 & -0.8 (0.78) & 0.70 & 0.96 \\

 $\beta_{2}$=1 & 1.8 (1.94) & 2.06 & 0.94 & -2.4 (2.61) & 2.59 & 0.95 & 6.3 (3.08) & 2.72 & 0.96 \\

 $\beta_{3}$=1 & 2.8 (0.57)  & 0.56 & 0.95 & 2.4 (0.66) & 0.65  & 0.95 & 1.0 (0.72) & 0.64 & 0.97 \\

$\beta_{4}$=1 & -1.1 (1.10)  & 1.20 & 0.93 & -4.1 (1.48) & 1.48  & 0.94 & -7.9 (1.76) & 1.53 & 0.96 \\

\bottomrule
\end{tabular}%
\end{adjustbox}
\label{cate_j=36}%
\end{table}%

Table \ref{cate_j=36} summarizes the results for the small sample simulation. The CDML estimates are again all close to simulated values with bias $<$ 3\% and small ASEs close to ESEs with good coverages near nominal 0.95. GSExact estimates are reasonably close to CDML estimates exhibiting small biases up to 4.1\% except a comparatively larger bias -7.4\% in the estimates of the level-2 variance $\tau$ from small sample sizes and high missing rates.

Blimp estimates exhibits larger biases than do GSEact estimates overall. Specifically, the biases in the estimated main effect $\beta_2$ of the binary covariate $D_j$, and the estimated interaction effect $\beta_4$ of $C_jD_j$ are 6.3\% and -7.9\% approximately twice as large as -2.4\% and -4.1\% produced by GSExact, respectively. The $\tau$ estimates are biased 18.9\% upwards by a larger magnitude than -7.4\% by GSExact. Other estimates by both approaches are comparatively accurate with biases lower than 3\%. Except for somewhat inflated ASE 3.14 associated with the $\tau$ estimates by Blimp compared to the GSExact counterpart 2.23, other ASEs, ESEs and coverages by Blimp and GSExact appear comparable. Observed small-sample differences in some biases by GSExact and Blimp result mainly because of different ways each method imputes missing values of $C_j$ and $D_j$: GSExact is based on exact posterior distributions of $C_j$ and $D_j$ while Blimp imputes $C_j$ and the underlying latent variable $D^*_j$ of $D_j$ by a Metropolis algorithm using normal proposal densities with constant variances.

\subsection{Robustness}
Because we simulated data from multivariate normality (\ref{eq:4covariate}) of continuous $C_j$ given binary $D_j$ as a part of the Bayesian joint model assumption of GSExact while Blimp estimation is based on multivariate normality of $C_j$ and the latent normal random variable $D^*_j$ underlying binary $D_j$, the simulation may be viewed to give GSExact an advantage to produce better estimates. To ensure a fair comparison with Blimp estimates and to assess the robustness of our estimators against a potential violation of the normality assumption of covariates, we conducted an additional simulation study.

Instead of simulating $C_j$ from model (\ref{eq:4covariate}), we simulate the data from the covariate model assumption of Blimp:
\begin{align} \label{eq:4blimpcovariate}
f(C_{1j}, D^*_j|C_{2j}) \sim N\left[ W\alpha = \left(I_{2} \otimes [1\; C_{2j}]\right) \alpha, T\right]
\end{align}
\noindent where the threshold parameter $\kappa$ determines $D_j=1$ if $D^*_j > \kappa$ and $D_j=0$ otherwise (\citealp{agresti2012categorical}; \citealp{carpenterandkenward}; \citealp{enders2018fully}; \citealp{enders2020model}). Because model (\ref{eq:4blimpcovariate}) belongs to the Bayesian joint distribution that is correctly assumed by Blimp but different from that of GSExact, Blimp now has an advantage to produce better estimates than does GSExact. 

We simulate sequentially: 1) $C_{2j}\sim N(2,1)$; 2) $C_{1j}\sim N(0.75+0.7X_j, 1.25)$ and $ D^*_{j} \sim N(-0.5+X_j, 1)$ having $T_{12}=cov(C_{j},D^*_{j})=-0.5$ in model (\ref{eq:4blimpcovariate}); 3) $D_{j}=1$ if $D^*_j > 2.2$, $D_{j}=0$ otherwise to set $p(D_{j}=1)=0.3$; and 4) $Y_{ij}\sim N(1+C_{1j}+D_{j}+C_{2j}+C_{1j}D_{j}, \tau+\sigma^2)$ for $\tau=4$ and $\sigma^2=16$ in HLM (1). Therefore, GSExact's bivariate normality assumption of $C_j$ conditional on $D_j$ violates the data generating mechanism of this simulation model. The simulated coefficients of the HLM all equal to 1 to facilitate the comparison again. We simulated 1000 data sets for the large and small sample cases, simulating the missing rates by Equation (\ref{eq:4missingrate}) as before.

Table \ref{robocatej=200} lists the estimated HLM for the large sample case. ALL CDML estimates are very accurate with small biases and ASEs and good coverages near nominal 0.95. Both GSExact and Blimp produce estimates close to the CDML estimates with small biases less than 2\% except for the main effect $\beta_2$ of $D_j$ and level-2 varaince $\tau$. The biases associated with the $\tau$ estimates are -3.4\% by GSExact and 3.4\% by Blimp with the same magnitude while the biases associated with the main effect $\beta_2$ of binary $D_j$ is 2.4\% by Blimp more accurate than -4.7\% by GSExact. Consequentely, Blimp produces a noticeably more accurate $\beta_2$ estimate than does GSExact, other estimates are comparably accurate by both approaches. ASEs are close to ESEs with coverage probabilities near nominal 0.95 by both approaches. The noticeable but modest difference between biases of $\beta_2$ estimates by the methods reflects the effect of a incorrectly assumed covariate model (\ref{eq:4covariate}) by GSExact. Overall, GSExact produce estimates quite robust against the incorrect model assumption in our large sample simulation scenario. 

\begin{table}[htbp]
  \caption{Estimated biases, ASEs, ESEs and coverages from the large sample simulation $(n_j=4, J=200)$.}
\begin{adjustbox}{max width=\textwidth}
\renewcommand{\arraystretch}{1.5}
    \begin{tabular}{cccccccccc}
\toprule
  & \multicolumn{3}{c}{CDML} & \multicolumn{3}{c}{GSExact} & \multicolumn{3}{c}{Blimp} \\
\cmidrule(lr){2-4}\cmidrule(lr){5-7}\cmidrule(lr){8-10}
 Simulated & \%Bias(ASE) & ESE & Coverage & \%Bias(ASE) & ESE & Coverage  & \%Bias(ASE) & ESE & Coverage \\ \midrule

 $\tau$=4 & 1.0 (-) & 0.91 & -& -3.4 (1.01) & 1.06  & 0.92 & 3.4 (1.06) & 1.08 & 0.93 \\

 $\sigma^2$=16 & 0.1 (-) & 0.91 & - & 0.8 (1.08) & 1.07 & 0.94 & 0.6 (1.08) & 1.06  & 0.94  \\
 
 $\beta_{0}$=1 & -0.1 (0.52) & 0.52 & 0.95 & -1.9 (0.58) & 0.59 & 0.95 & -0.1 (0.58) & 0.58 & 0.96 \\

 $\beta_{1}$=1 & -0.1 (0.22) & 0.21 & 0.95 & -1.4 (0.26) & 0.25 & 0.96 & -0.2 (0.26) & 0.25 & 0.96 \\

 $\beta_{2}$=1 & -0.5 (0.94) & 0.92 & 0.95 & -4.7 (1.10) & 1.10 & 0.95 & 2.4 (1.13) & 1.10 & 0.96 \\

 $\beta_{3}$=1 & -0.1 (0.30)  & 0.28 & 0.96 & 0.2 (0.34) & 0.33  & 0.95 & -0.1 (0.34) & 0.33 & 0.95 \\

 $\beta_{4}$=1 & -0.6 (0.34)  & 0.34 & 0.95 & -0.2 (0.40) & 0.40  & 0.95 & -1.2 (0.40) & 0.40 & 0.96 \\
 
\bottomrule
\end{tabular}%
\end{adjustbox}
\label{robocatej=200}%
\end{table}%

\begin{table}[htbp]
  \caption{Estimated biases, ASEs, ESEs and coverages from the small sample simulation $(n_j=4, J=36)$.}
\begin{adjustbox}{max width=\textwidth}
\renewcommand{\arraystretch}{1.5}
    \begin{tabular}{cccccccccc}
\toprule
  & \multicolumn{3}{c}{CDML} & \multicolumn{3}{c}{GSExact} & \multicolumn{3}{c}{Blimp} \\
\cmidrule(lr){2-4}\cmidrule(lr){5-7}\cmidrule(lr){8-10}
 Simulated & \%Bias(ASE) & ESE & Coverage & \%Bias(ASE) & ESE & Coverage  & \%Bias(ASE) & ESE & Coverage \\ \midrule

 $\tau$=4 & -0.1 (-) & 2.17 & -& -5.5 (2.32) & 1.84  & 0.97 & 23.2 (3.28) & 2.51 & 0.97 \\

 $\sigma^2$=16 & 0.4 (-) & 2.20 & - & 2.4 (2.57) & 2.46 & 0.95 & 2.9 (2.66) & 2.47  & 0.95  \\
 
 $\beta_{0}$=1 & -4.5 (1.31) & 1.36 & 0.94 & -9.1 (1.53) & 1.50 & 0.94 & -7.9 (1.66) & 1.48 & 0.96 \\

 $\beta_{1}$=1 & 0.2 (0.54) & 0.59 & 0.93 & -3.6 (0.69) & 0.70 & 0.94 & -4.8 (0.73) & 0.68 & 0.96 \\

 $\beta_{2}$=1 & 0.2 (2.45) & 2.72 & 0.92 & -9.5 (3.18) & 3.14 & 0.94 & -6.3 (3.60) & 3.20 & 0.96 \\

 $\beta_{3}$=1 & 2.9 (0.75)  & 0.77 & 0.94 & 8.9 (0.89) & 0.92  & 0.94 & 12.9 (0.97) & 0.91 & 0.96 \\

$\beta_{4}$=1 & -2.9 (0.91)  & 1.02 & 0.92 & -3.3 (1.19) & 1.18  & 0.93 & -4.9 (1.37) & 1.20 & 0.95 \\

\bottomrule
\end{tabular}%
\end{adjustbox}
\label{robocate_j=36}%
\end{table}%

Table \ref{robocate_j=36} summarizes the simulation results from the small sample scenario. Compared to the results from a large sample simulation, the CDML estimates result in noticeably higher biases up to -4.5\% in magnitude but reasonable accuracy overall, and the associated standard errors increase due to the small sample effect, too. ASEs are close to ESEs with good coverage near nominal 0.95. GSExact estimates of the intercept $\beta_0$ and the main effects $\beta_2$ of $D_j$ and $\beta_3$ of $C_{2j}$ are -9.1\%, -9.5\% and 8.9\% biased respectively, and the level-2 variance $\tau$ estimate is -5.5\% biased while other estimates are accurate with biases well below 5\% in magnitude. At least some of the biases above 5\% should reflect the incorrect covariate model assumption by GSExact. ASEs are modestly inflated compared with the CDML counterparts due to extra uncertainty from missing data, but close representations of ESEs. Coverages are close to nominal 0.95. 

Although Blimp estimation has a comparative advantage based on the correctly specified latent covariate model (\ref{eq:4blimpcovariate}), some estimates are less accurate that GSExact estimates. Specifically, Blimp biases 23.2, 2.9, -4.8, 12.9 and -4.9\% in the estimates of $\tau, \sigma^2, \beta_1, \beta_3$ and $\beta_4$ are higher than GSExact counterparts -5.5, 2.4, -3.6, 8.9 and -3.3\%, respectively, while Blimp biases -7.9\% and -6.3\% in the estimates of $\beta_0$ and $\beta_2$ are lower than GSExact counterparts -9.1 and -9.5\%, respectively. Blimp ASEs are slightly higher than ESEs and GSExact ASEs overall, but coverages are near the nominal level. Although GSExact assumes an incorrect covariate model assumption (\ref{eq:4covariate}), GSExact estimates appear comparatively robust by producing estimates slightly more accurate and precise than Blimp estimates overall under this small sample scenario.  

\section{Analysis of Racially Discordant Patient-Physician Interactions}
We use GSExact to analyze data from the study of racially discordant medical interactions between patients and physicians. Investigators videotaped and coded physicians' communication behavior and facial expression while the physicians interacted with patients during office visits. The study aimed to understand how physician's implicit prejudice and patient income are associated with the way that racially discordant physicians communicate with Black patients.

Each physician completed a baseline survey on demographics and other characteristics including implicit prejudice ($IPrej$) measured by the Implicit Association Test \citep{greenwald1998measuring,nosek2005understanding} and time elapsed since the last communication training (CT) that is designed to improve communication skills with patients. Patients also completed a survey before the office visit on demographics including their age, gender, race, education level, income as well as patient characteristics related to their medical statuses. Subsequent medical encounters between physicians and patients were videotaped during office visits. The video-taped facial expression and communication behavioral of a physician was rated as a positive valence score by a machine on four consecutive time points during a 20-minute encounter. A positive valence score measures the intensity of positive facial expression. Due to COVID-19 restriction, investigators were able to recruit only 37 patients and 6 physicians much less than originally planned. 

We estimate the main and interaction effects of physician's implicit prejudice and patient's family income on the valence score outcome during the medical encounters. Income level is 1 if the family income of a patient is below \$25,000 and 0 otherwise. We also analyze a completely observed physician's communication training (CT) covariate. To improve the interpretability of the model, we center covariates at their sample means. Since the positive valence score tends to be higher at the beginning and end (e.g., when greeting) than mid time points of the medical encounter (e.g., when talking about serious medical conditions), we also control for dummy variables indicating time $Q_2$, $Q_3$ and $Q_4$ indicating time points 2, 3 and 4, respectively. We write our HLM:
\begin{align} \label{eq:4realdataHLM}
Valence_{ij}=\beta_0+\beta_1IPrej_j+\beta_2Income_j+\beta_3CT_j+\beta_4IPrej_jIncome_j+\beta_5Q_{2ij}+\beta_6Q_{3ij}+\beta_7Q_{4ij}+u_j+e_{ij}
\end{align}
\noindent where occasion $i$ is nested within the $j$th patient-physician encounter for $i=1,\cdots,4$ and $j=1,\cdots,37$. Here  $\beta_0$ is the expected valence of the encounter between a physician having the average implicit prejudice ($IPrej$), average level of communication training ($CT$) and a patient with a high family income at occasion 1 or $Q_1$. Fixed effects are the main effects $\beta_1$, $\beta_2$ and $\beta_3$ of $IPrej$, $Income$ and $CT$, respectively; the interaction effect $\beta_4$ of $IPrej \times Income$; and the mean differences $\beta_5-\beta_7$ in the outcome at occasions 2-4 relative to occasion 1, ceteris paribus. Encounter-specific random effect $u_j\sim \mathcal{N}(0,\tau)$ and patient-specific random effect $e_{ij} \sim N(0,\sigma^2)$ are independent. Valence score is missing 20\%, $IPrej$ is missing 16\% and income is missing 11\% of the values.

\begin{table}[htbp]
\centering
  \caption{Estimated HLM (\ref{eq:4realdataHLM})}
\renewcommand{\arraystretch}{1.5}
    \begin{tabular}{cccc}
        \toprule
 Parameter & Covariate & Estimate(SE\textsuperscript{+}) & CI\textsuperscript{++}(2.5th, 97.5th)  \\ \midrule
 
 $\beta_{0}$ & Intercept & 88.82(2.98)\textsuperscript{*} & (82.85, 94.40)    \\

 $\beta_{1}$ & IPrej &   6.26(6.69) & (-6.83, 19.19)  \\

 $\beta_{2}$ & Income & -7.63(2.77)\textsuperscript{*}  & (-12.93, -2.07)    \\

 $\beta_{3}$ & CT &   1.52(1.68)  & (-1.75, 4.92)  \\
 
 $\beta_{4}$ & IPrej$\times$Income & -20.36(9.52)\textsuperscript{*}  & (-38.98, -2.06)   \\
 
 $\beta_{5}$ & Q2 &  -9.33(2.48)\textsuperscript{*}  & (-14.29, -4.29)   \\

 $\beta_{6}$ & Q3 &  -4.24(2.57)  & (-9.39, 0.48)   \\

 $\beta_{7}$ & Q4 &  -1.97(2.39)  & (-6.62, 2.86)   \\

 $\tau$      & - &   3.48(3.79) & (0.49, 14.48)   \\

 $\sigma^2$  & - &  77.81(11.59) & (58.18, 103.33)  \\

\bottomrule
\multicolumn{4}{l}{\footnotesize{+:standard error; ++:a 95\% credible interval; *:significantly different from zero at a level 0.05}}
\end{tabular}%
\label{realdata_cate}%
\end{table}%

Table \ref{realdata_cate} shows the estimated HLM by GSexact. Each estimate and its associated standard error in parenthesis appear in column three followed by a 95\% Bayesian credible interval in the last column. The estimated main effects $\beta_2$ and $\beta_5$, and interaction effect $\beta_4$ are significantly different from zero at a level 0.05 as they lie outside the respective 95\% Bayesian credible intervals. Income (-7.63 (2.77)) and its interaction with implicit prejudice (-20.36 (9.52)) are negatively associated with the valence score (-7.63 (2.77) and the mean valence at occasion 2 is significantly lower than the mean outcome at occasion 1 (-9.33 (2.48)), controlling for other covariates in the model. Consequently, a low-income patient encountering a physician with average implicit prejudice is expected to encounter more unfriendly communication behavior than a high income patient meeting the physician (-7.63 (2.77)), and confront much more unfriendly communication behavior when meeting a physician with above-average implicit prejudice, ceteris paribus. The intra-cluster correlation coefficient 3.48/(3.48+77.81)=0.04 reveals that comparatively small 4\% of the total variance in valence remains to be explained at the encounter level. Lastly, each PSRF near or less than 1.1 implies the Gibbs sampler met convergence criteria.

\section{Discussion}

We estimated a hierarchical linear model (HLM) from incompletely observed longitudinal data nested within the small number of patient-physician encounters during office visits that were assumed missing at random (MAR). A mixture of correlated categorical and continuous covariates had interaction effects on a continuous outcome, and the interactive covariates, as well as the outcome, may be partially observed. We obtained efficient and unbiased estimation of the model by a Gibbs sampler that is not only based on exact posterior distributions, but also guaranteed to be compatible with the HLM. Our simulation results showed that the sampler produced reasonably accurate, precise and robust estimates, even in situations where the sample sizes are as small as 4 units nested within 36 clusters. Our estimates were as accurate and precise as those of competing methods given large sample sizes (4 units nested within 200 clusters), and more accurate and precise overall than competing ones given the small sample sizes.

To handle missing data efficiently, we assumed normally distributed continuous covariates conditional on categorical covariates and dependent categorical covariates from a multinomial distribution. A valuable future research topic is to extend the exact Gibbs sampler approach to incompletely observed interactive covariates and outcome from other exponential family distributions such as count, ordinal categorical and non-normal continuous variables. Because our simulation study focused on the same cluster size fixed at $n=4$ as our motivating example while the number of $J$ clusters was varied, another valuable future research area is to evaluate our methods as cluster sizes vary. Finally we plan to extend our sampler to efficiently estimate a HLM with the interaction effects of partially observed lower-level covariates.

\nocite{*}
\clearpage
\bibliographystyle{apalike}

\clearpage
\end{document}